# Effect of cathode porosity on the Lithium air cell oxygen reduction reaction – a rotating ring-disk electrode investigation

Jeongwook Seo[a,*,#] Shrihari Sankarasubramanian[a,*,#,1], Nikhilendra Singh[b], Fuminori Mizuno[b,c], Kensuke Takechi[b], and Jai Prakash[a,*]

[a] *Center for Electrochemical Science and Engineering, Department of Chemical and Biological Engineering, Illinois Institute of Technology, 10 West 33rd Street, Chicago, IL 60616, USA*

[b] *Toyota Research Institute of North America, 1555 Woodridge Avenue, Ann Arbor, Michigan 48105, USA*

## Abstract

The kinetics of the oxygen reduction reaction (ORR) on the practical air cathode in a Lithium air cell, which is conventionally composed of porous carbon with or without catalysts supported on it, was investigated. The mechanism and kinetics of the oxygen reduction reaction (ORR) was studied on a porous carbon electrode in an oxygen saturated solution of 0.1M Lithium bis-trifluoromethanesulfonimidate (LiTFSI) in Dimethoxyethane (DME) using cyclic voltammetery (CV) and the rotating ring-disk electrode (RRDE) technique. The oxygen reduction and evolution reactions were found to occur at similar potentials to those observed on a smooth, planar glassy carbon (GC) electrode. The effect of the porosity and the resultant increase in surface area were readily observed in the increase in the transient time required for the intermediates to reach the ring and the much larger disk currents (compared to smooth, planar GC) recorded respectively. The RRDE data was analyzed using a kinetic model previously developed by us and the rate constant of the elementary reactions calculated. The rates constant for the electrochemical

[*] To whom correspondence may be addressed: E-mail address: jseo16@hawk.iit.edu (J.Seo), ssanka11@hawk.iit.edu (S, Sankarasubramanian), prakash@iit.edu (J.Prakash)

[#] These authors contributed equally to this work.

[c] Present address: Battery Material Engineering & Research Division, Toyota Motor Corporation, Higashifuji Technical Center, 1200, Mishuku, Susono, Shizuoka, 410-1193 Japan

[1] Present address: Department of Energy, Environmental and Chemical Engineering, Washington University in St. Louis, 1 Brookings Dr, St. Louis, MO 63130, USA

reactions were found to be similar in magnitude to the rate constants calculated for smooth GC disks. The porosity of the electrode was found to decrease the rate of desorption of the intermediate and the product and delay their diffusion by shifting it from a Fickian regime in the electrolyte bulk to the Knudsen regime in the film pores. Thus, it is shown that the effect of the electrode porosity on the kinetics of the ORR is physical rather than electrochemical.

1. **Introduction**

The mitigation of "range anxiety" hindering the adoption of electric vehicles (EV) requires that EVs on a single charge match the range of a gasoline powered car with a full tank. A range of 500 miles/charge, which would satisfy this requirement, requires the commercialization of "beyond Li-ion" battery systems. The Li-$O_2$ cell is one of the best candidates due to its high theoretical specific energy of 3505 Wh $Kg^{-1}$ which is significantly higher than current lithium-ion cells [1,2]. The practical use of this promising technology is beset by problems of the high overvoltage leading to low columbic efficiency and low power density [1-3], Li dendrite growth on the anode [4], side reactions with $N_2$ and $H_2O$ in air [5,6], and unstable electrolytes leading to short cycle life [7-10],

During oxygen reduction reaction (ORR), $O_2$ or $O_2^-$ formed by outer sphere electron transfer, is adsorbed onto the cathode and reacts with $Li^+$ to form $LiO_2$ though net one electron transfer. $Li_2O_2$ is formed from $LiO_2$ via a subsequent electrochemical reaction or by chemical disproportionation reaction [10, 11]. An examination of the literature indicates that the two-electron transfer reaction has not been reported. A consensus exists in the literature that $Li_2O_2$ is the final discharge product of the Li-$O_2$ cell following extensively studies using in-situ [12-14] and ex-situ techniques [15] but the elementary steps involved in the ORR are less well understood. The net one electron transfer observed herein has led to the chemical disproportionation of $LiO_2$ being considered the major reaction route [14-19] with reports of its occurrence in the electrolyte bulk [15-17], on the electrode surface [18,19] or as a hybrid surface-electrolyte mechanism [14]. It is evident from the literature that analyses of data from the same technique such as SERS [12,38], AFM [34,40] or EQCM [19,34] for example, have led to radically different conclusions as to the overall mechanism. Several of these studies [12,34,40] also employ unstable [11,41,42] DMSO as the electrolyte solvent, necessitating the data obtained to be examined in this context. Thus, the elucidation of the Li-O2 ORR mechanism is complicated by the interplay of factors such the solvent [15,16,21], salt [43,44], applied overpotential [14] and electrolyte water content [6,21].

The applicability of the rotating ring-disk electrode (RRDE) technique has been reported previously [11, 21, 22, 35] to elucidate complex reaction mechanisms in non-aqueous systems. The present study seeks to use the RRDE technique to examine the $Li^+$ ORR on practical, porous carbon electrodes. Porous carbon electrodes have been extensively used as cathodes for Li-air cells [3,4,10,22-27]. The influence of the carbon surface on $Li_2O_2$ deposits on the cathode has bene examined by us previously [23] and others have examined the corresponding oxygen evolution reaction (OER) [25]. In the present study, we combine RRDE measurements and a kinetic model to calculate the rate constants for each of the elementary steps and the competition between the surface and bulk reactions is quantified.

2. **Experimental Methods**

The carbon black mixture was prepared by mixing 10 mg Carbon black (CB, acetylene, 100% compressed, surface area: 80 $m^2g^{-1}$, >99.9 %, Alfa Aesar) with 10 wt% polytetrafluorethylene solution (PTFE, 60 wt% dispersion in water, Sigma Aldrich) followed by addition of 3 ml isopropanol to the CB/PTFE mixture solution to dilute the mixture. The mixture was dispersed with a sonicator (Qsonica, frequency: 20 KHz) for 10 minutes at room temperature. The glassy carbon (GC) disk of the RRDE system was roughened using sand paper and recessed ~ 1 mm below the lip of the surrounding teflon holder. The CB film was deposited by dropping 7 μl CB/PTFE slurry onto the recessed, roughened GC disk electrode (surface area: 0.196 $cm^2$) and dried in an oven at 40 °C for 15 minutes. This CB coating process was repeated three times until it uniform surface coverage was achieved and the film was flush with the RRDE assembly surface. Finally, the porous CB disk electrode was dried in a vacuum oven at 40 °C for 12 hours to remove all water and solvent.

The 0.1M solution of Lithium bis-trifluoromethanesulfonimidate (LiTFSI) (99.95%, Aldrich) in 1,2-Dimethoxyethane (DME) (Sigma-Aldrich, ReagentPlus, ≥99%) was prepared in a MBraun argon filled glove box with $H_2O$ and $O_2$ levels <0.5ppm. The LiTFSI salt was dried in a vacuum oven for 24 hours at 40°C and DME was distilled, argon purged and stored in an argon filled glove box to ensure absence of moisture before use. The water content in the "as-received" DME was measured using Karl-Fisher titration (Mettler Toledo C30 Coulometric KF Titrator) and the water content was found to be 77.6 ppm. Similarly, the water content of the electrolyte was measured after preparation and found to be 76.8 ppm.

The electrochemical experiments were carried out with a multi-channel potentiostat (Solartron Analytical) with independent leads for potential and current. A 50-ml flask was used for the RRDE measurement with a custom designed Teflon top with four openings for working, reference, counter electrode and the gas purge line respectively. The three-electrode system consisted of Pt gauze (Alfa Aesar, 45 mesh, <99.9%) on a Pt wire as reference and counter electrode. The working electrode was a Pine instruments RRDE with the CB coated GC disk surrounded by a Pt ring with a rated collection efficiency (N) of 0.25 and used in conjunction with the supplied rotor shaft and gas purged bearing assembly. The entire setup was assembled in the glove box and sealed. Upon transfer out of the glovebox, the gas inlet was immediately connected to an Ar source to maintain the Ar blanket over the electrolyte surface. The gas purged bearing assembly was connected to a second Ar source to prevent ingress of atmospheric air during RRDE shaft rotation. The Pt pseudo-reference was calibrated with respect to the $Li/Li^+$ couple using a ferrocene/ferrocenium ($Fc/Fc^+$) internal reference as recommended by IUPAC [45]. The calibration procedure and resultant graphs are in supplementary materials section 1.

The background cyclic voltammograms (CV) were measured in Ar saturated state after a 30 min Ar purge. The CVs were then measured using different scan rates in the $O_2$ saturated condition after $O_2$ purge for 1 hour. For the RRDE measurements, the working electrode rotation rate was controlled by a MSRX speed control (Pine Instrument). The ring potential was held at 3.08 V which was chosen to achieve complete oxidation of reduced species produced at the disk. The linear sweep voltammograms (LSVs) were measured from 2.2 to 1.0 V vs $Li/Li^+$ based on the location of the ORR peak identified from the CV.

## 3. Results and discussions

*3.1 Electrochemical measurements*

The oxygen reduction and oxygen evolution current was measured on the porous CB coated GC electrode disk. Figure 1 shows CVs performed over the potential range of the ORR and OER with different scan rates in $O_2$ saturated 0.1 M LiTFSI in DME electrolyte. The background CV was measured in Ar saturated electrolyte and parasitic side reactions were not observed in this voltage range. A peak seperation of 1300 mV was measured with 5 mV.s$^{-1}$ scan rate and hence was found to be higher than the theoretical value for a one electron reversible reaction (59 mV). The ORR and OER peak shifts were seen to be in agreement with the suggestion of a 30/α mV shift for a 10 fold increase in scan rate [29].

The electrochemical kinetic analysis for ORR was carried by combining the RRDE technique with an electro-kinetic model. A detailed discussion of the use of RRDE measurements for non-aqueous ORR reactions can be found in our recent works [11, 20]. Figure 2 shows the RRDE measurements on the porous CB disk and Pt ring system carried out at a scan rate of 2 mV.s$^{-1}$. The LSVs of porous CB disk showed the typical observation that the limiting currents ($i_L$) is directly proportional to the square root of rotation rate ($\omega^{-\frac{1}{2}}$).

*3.2 Kinetic analysis*

The Koutecky-Levich equation describes the current measured by a RRDE if the ORR kinetics is first order with respect to dissolved oxygen and the disk current and square root of rotation rate have the relation given by equation (1):

$$\frac{1}{i} = \frac{1}{i_k} + \frac{1}{i_L} = \frac{1}{i_k} + \frac{1}{B\sqrt{\omega}} \tag{1}$$

where $i_k$ is the kinetic current. Thus, the slope of a typical Koutecky-Levich (K-L) plot ($i^{-1}$ vs. $\omega^{-\frac{1}{2}}$) as shown in Figure 3 would equal a constant B given by equation (2):

$$B = 0.62\, nFAD^{\frac{2}{3}}v^{-\frac{1}{6}}C_b \tag{2}$$

Where $F$ is 96,485 C mol$^{-1}$, $A$ is the geometric area of disk, $D$ is the diffusion coefficient of O$_2$ (2.78 x 10$^{-6}$ cm$^2$s$^{-1}$ in 0.1M LITFSI/DME) [22], $Cb$ is the bulk concentration of O$_2$ (9.57 x 10-6 mol/cm$^3$) [30], ν is kinematic viscosity (5.22 x 10$^{-3}$ cm s$^{-1}$) that was measured using a Ubbelohde viscometer. The presence of the CB film potentially complicates the analysis of the K-L plot compared to the case of the smooth electrode described above. In case the film was porous enough to allow for flow inside the pores, the current versus rotation rate profile has been shown to have a sigmoidal profile with the current being linear below a certain critical rotation rate and increasing exponentially above this value [33]. On the other hand, for dense porous materials like most carbons used in electrodes, the pores would be too small for flow to occur inside them [33]. The transport regime in the film in such cases is expected to be dominated by diffusion, with the contributions from Fickian and Knudsen diffusion depending on the pore size distribution. Apart from the convective transport to the disk surface and kinetics on the disk surface that can serve as the limiting factor (depending on the potential), in a porous electrode system we may also need to consider the diffusive transport through the film, possible reaction rate limitation at the redox centers inside the film and partitioning between the film surface and the film bulk (i.e. inside the pores) [29]. The various possibilities and the Koutecky-Levich like equations applicable to these cases are comprehensively discussed elsewhere [29,36]. It is to be noted that the effect of these various possible limiting factors is typically an additional contribution to the limiting current leading to a non-zero intercept for the *1/i* vs. $\omega^{-\frac{1}{2}}$ plots at the limiting current region. Depending on the film and the postulated limiting factor in the film, this non-zero intercept is treated suitably. The slope of the K-L plot and hence kinetic calculations such as the number of electrons is unaffected. In the present case, for a uniform, conducting carbon film whose thickness is an order of magnitude less than its diameter, the diffusion process in the film is believed to be dominant contribution to the film current. To account for this film current, we may re-write the K-L equation as follows [29]:

$$\frac{1}{i} = \frac{1}{i_k} + \left[\frac{1}{i_{Levich}} + \frac{1}{i_{film}}\right] \quad (3)$$

The depletion layer at the disk disappears at high enough rotation rate and hence we may assume that the reactant concentration exposed to the disk would be the same as the bulk concentration. Thus, the film current maybe represented as [29]:

$$i_{film} = \frac{nFAD_{film}\kappa C_{O_2,bulk}}{\phi} \quad (4)$$

The reaction occurring on the film surface vs. the pores can be distinguished through the partition coefficient $\kappa$ (assuming the same rate expression in both the pores and the surface) and the pore diffusion (possibly Fickian) can be characterized by the film diffusion coefficient $D_{film}$. Interestingly, it was observed that while there is a small film current at the start of the limiting current region (1.8V), this current disappears at higher overpotentials (1.4V). This may indicate that at high enough overpotentials the reactions on the surface are so rapid as to consume all reactants reaching the surface before they have an opportunity to diffuse into the film bulk. Nevertheless, calculating the $\frac{D_{film}\kappa}{\phi}$ term using the film current at 1.8V (1.03mA), a value of 5.7 x $10^{-5}$ m.s$^{-1}$ was obtained. Since the film is ~1mm thick, this leads to a film diffusion coefficient of 5.7 x $10^{-8}$ m$^2$. s$^{-1}$. This is two orders of magnitude lower than the bulk diffusion coefficient of oxygen (2.78 x $10^{-6}$ cm$^2$s$^{-1}$ in 0.1M LITFSI/DME).

Alternatively, the contribution of film diffusion can be evaluated using the disk to ring transient time. The transient time is related to the reactant diffusion coefficient as follows [37]:

$$T_s = K \left(\frac{\nu}{D_{O_2,bulk}}\right)^{\frac{1}{3}} \omega^{-1} \qquad (5)$$

K is a RRDE geometry dependent proportionality factor given by $K = 43.12 \left(\log\left(\frac{r_2}{r_1}\right)\right)^{\frac{2}{3}}$. In the present case, it is 10.1 rpm.s, leading to a calculated value of the transient time at 400 rpm of 0.3s. Whereas experimentally, we observe a transient time of ~14s. While there may be a contribution from the difference between the calculated and actual K values (due to variations in RRDE geometry) these would not be sufficient to explain the two orders of magnitude variation (the experimentally determined value of K in reference 37 was 11). But since the film and bulk diffusion coefficients vary by two orders of magnitude, the possible reason for this discrepancy is apparent. The slow product diffusion through the film on the disk, slows down its detection at the ring. Thus, analysis of both the reaction currents at the disk and product oxidation currents at the ring show that transport through the film plays a major role in the determining reaction site (film surface vs. film pores) and rate of product transport through the film ($Li_2O_2$ precipitation vs. pore clogging). This is a vital first step in understanding the $Li^+$ ORR in practical electrodes.

The plots from the mixed control region between 1.5 V to 1.7 V were used to calculate the total number of electrons transferred in the reaction and this was found to equal 1.07 within 10% error. This calculation was carried out using the geometric surface area of the electrode. The actual surface area measured by techniques such as $N_2$ adsorption (BET) is not applicable herein as the electroactive surface area is not necessarily the same as the physical surface area. The variation of ORR kinetics by adsorption site has been discussed by us elsewhere [20]. Further, the bulk diffusion coefficient of oxygen is used as the effective diffusion coefficient.

The combination of the total number of electrons with the information on the rate determining step obtained from a Tafel plot allowed us to describe the overall electrochemistry of the present system. Semilog plots of kinetic current and the disk potential (Tafel plot) are depicted in the Figure 4. The kinetic current and the overpotential is related by the Tafel equation:

$$\eta = a + b.\log(i_k) \qquad (6)$$

Where

$$a = \frac{2.3RT}{\alpha.F}.\log i_0 \qquad (7)$$

$$b = -\frac{2.3RT}{\alpha.F} \qquad (8)$$

The kinetic current $i_k$ is determined by a rearrangement of the Koutecky-Levich equation ($i_k = i.i_L/(i_L - i)$) to obtain a relationship between the disk current and the limiting current. The slope of this plot indicates the rate determining step. In the present case, a value of 390 mV.dec$^{-1}$ was obtained which was found to be at odds with the theoretical value (118 mV.dec$^{-1}$ with transfer coefficient ($\alpha$) = 0.5) for the first electron transfer step being rate determining. This anomalous value of the Tafel slope points to deviation in the value of the transfer coefficient from the conventional value of 0.5 to 0.16 as described by us previously [22]. This indicates an asymmetric energy landscape for the anodic and cathodic directions of the cathode reactions caused by the solvent interactions has been described by us elsewhere [22].

To calculate the individual reaction rate constants for the elementary steps multiple possible Li-O$_2$ ORR pathways depicted in the Figure 5 were considered. The initial step consists of the adsorption of O$_2$ on the cathode surface and subsequent production of LiO$_2$. Following the production of LiO$_2$, various possible reduction pathways were considered – further electrochemical reduction to produce Li$_2$O$_2$, chemical disproportionation reaction on the surface between adjacent LiO$_2$ to produce Li$_2$O$_2$ and desorption followed by chemical disproportionation to produce Li$_2$O$_2$ in the electrolyte bulk. However, the four-electron reaction to produce Li$_2$O was not considered in this reaction scheme due to lack of experimental reports of Li$_2$O as the reduction product. The effect of water on the reaction was a key consideration. It has been reported that the surface or the solution phase reaction dominates based on the water content [46]. The present model considers both surface and solution mechanisms and hence this variability is implicitly accounted for. Further, the amount of water is in between the limits for transition between a largely surface based (<30ppm) to a largely solution based mechanism (> ~2000ppm). Hence a model based on competitive solution and surface based mechanisms would be most representative of the present system. Efforts to explicitly incorporate side reactions with water, its possible catalytic activity and its effects on the nucleation of Li$_2$O$_2$ are ongoing.

The reactions were described by typical kinetic equations along with equations for the disk and ring currents. The equations are as follows:

$$k_1[O_{2(a)}] = (k_3 + k_4 + k_5)[LiO_{2(a)}] \tag{9}$$

$$k_5[LiO_{2(a)}] - k_6[LiO_{2*}] = Z\omega^{\frac{1}{2}}[LiO_{2*}] \tag{10}$$

$$i_D = nFA\{(k_1 + 2k_2)[O_{2(a)}] + k_3[LiO_{2(a)}]\} \tag{11}$$

$$i_R = nFANZ\omega^{\frac{1}{2}}[LiO_{2*}] \tag{12}$$

Where Z is a mass transport relation given by:

$$Z = 0.62 D^{\frac{2}{3}} \nu^{-\frac{1}{6}} \tag{13}$$

RRDE kinetics models from the literature for aqueous ORR [31, 32] and our prior work [11, 21, 22] were adapted to the present system to obtain the following relating the disk and ring current:

$$N\left(\frac{I_D}{I_R}\right) = \left(\frac{Ak_6}{k_5}\right)\left(\frac{1}{Z\omega^{\frac{1}{2}}}\right) + \left(\frac{A}{k_5}\right) \tag{14}$$

Where A is reaction rate constant relation given by:

$$A = 1 + \left(\frac{2k_2(k_3 + k_4 + k_5)}{k_1}\right) + k_3 \tag{15}$$

The kinetic relations were simplified based on our experimental observations. The Koutecky-Levich plots indicated that the overall reaction involved the transfer of a single electron. Thus, the direct two-electron and series based two-electron pathway to produce Li$_2$O$_2$ was eliminated ($k_2 = k_3 = 0$). The simplified model was further analyzed to obtain individual reaction rate constants.

Figure 6 shows the distribution of disk and ring current ($N(i_D/i_R)$) vs. square root of the rotation rate ($\omega^{-1/2}$). This plot is expected to yield straight lines with different slopes ($S$) and intercept ($J$) at different potentials based on the simplified form of equation (14) where A=1. The slopes ($S$) and intercepts ($J$) at different potentials from Figure 6 were plotted in the Figure 7. The slope and intercept are related by the following equation:

$$J = \left(\frac{Z}{k_6}\right) S \tag{16}$$

Figure 7 was seen to follow the prediction from the model that the intercept is zero and slope of the above plot allowed us to calculate the rate constant for the electrolyte bulk disproportionation reaction ($k_6$). The rate constant for desorption of LiO$_2$ from the surface ($k_5$) was then calculated from the slopes of the $N\left(\frac{i_D}{i_R}\right)$ plots. The value of $k_5$ was found to be smaller by three orders of magnitude as compared to the values the smooth, planar GC disk [22]. The rate of desorption was understood to be a function of the convective transport from the film surface reaction sites and diffusive (Fickian and/or Knudsen) transport from the intra-pore reaction sites. The material dependent intermediate binding energy could also play a role. The carbonaceous nature of the electrode in both the smooth and porous cases and the equal rates of rotation indicated that the variation in the desorption rates was a function of the porosity of the electrode. At the same rotation rates (i.e. equal rates of convection), the smooth electrode would display purely Fickian diffusion whereas the porous electrodes can be expected to display a mixture of Fickian and Knudsen behavior (which is typically slower) depending on pore size distribution. Further, desorption would also depend on the species concentration gradient between the surface and the near surface electrolyte environment. This gradient could be non-uniform in highly tortuous electrodes and hence lead to non-uniform mass transport driving forces leading to a net slower rate of desorption. This was reflected in the large time transient between the observation of the mass transport limited current at the disk and the appearance of the corresponding current plateau at the ring.

The rate constant for the first electron transfer ($k_1$) was calculated from the rotation dependent relation between $i_D$, $i_L$ described by equation (17) and depicted in Figure 8. The close match between the theoretical and experimental intercepts served as further validation of the present model.

$$\frac{i_L}{i_L - i_d} = 1 + \frac{k_1}{Z} \omega^{-\frac{1}{2}} \qquad (17)$$

Having calculated $k_1$, $k_5$ and $k_6$, $k_4$ was calculated by rearranging equations (9)-(12) as follows:

$$k_4 = \frac{nFAZ_{O_2}\omega^{\frac{1}{2}}[O_{2(b)}]}{i_D}\left(\frac{k_1}{1+k_1}\right) - k_5 \qquad (18)$$

The potential dependence of the various rate constants is summarized in Figure 9 and compared to the values for the smooth GC disk case from Sankarasubramanian et al. [22]. The potential dependent rate constant for the electrochemical one-electron transfer reaction ($k_1$) for both porous CB and smooth GC disk showed similar values as expected for chemically similar carbonaceous surfaces. This clearly indicated that any observed variation in the Li-$O_2$ cathode reactions with changes in electrode porosity is physical in origin and not chemical. The rate constants for the chemical disproportionation reaction in the bulk ($k_6$) was found to be similar for both electrodes since the reactions in the electrolyte are independent of the physical or chemical properties of the electrode. The non-electrochemical steps - chemical disproportionation on the surface and desorption of $LiO_2$ and their associated rate constants $k_4$ and $k_5$ were found to display no potential dependence as expected. The value of $k_4$ was found to slightly increase with potential in the case of the porous CB disk. This was understood to occur due to a combination of the increasing production of $LiO_2$ at higher potentials and the hindered desorption of the $LiO_2$ due to the porous nature of the electrode and its associated transport limitations. The surface disproportionation reaction required the presence of two $LiO_2$ species in proximity and the occurrence of this increases with increasing concentration of the superoxide. While on the smooth electrode this increase in concentration can be expected to result in a proportionate increase in the rate of desorption, this was not understood to be the case on the porous electrode as the effective rate of pore and bulk transport was previously seen to be two orders of magnitude lower. Thus, the smooth GC electrode showed a slight decrease in the rate of surface disproportionation with potential while the porous electrode was found to display the opposite behavior.

Thus, the porous electrode increases the retention of the product species on the electrode and hence offers a possible avenue towards greater reversibility of $Li_2O_2$, with the trade-off being the overall slower rate of reaction due to the slowdown in the rate of the chemical disproportionation reaction on the surface. Further, the chemical disproportionation reaction is expected to dominate due to the transport properties of a porous electrode.

**4. Conclusion**

The present study examines the effect of porosity on the individual reactions comprising the ORR in Li-$O_2$ systems. Porous carbon was chosen as being representative of the typical electrode and catalyst support used in Li-$O_2$ cells. The porosity of the electrode was found to have a significant impact on the desorption of reaction intermediates and their further reaction on the surface while having minimal effect on the electrochemical reaction or chemical reactions in the electrolyte bulk. Porous electrodes offer a way to increase reversibility of the Li-O2 cell by minimizing loss of the $Li_2O_2$ product by precipitation but would also be hampered by increased passivation. The present study is expected to inform the choice of porous materials as non-aqueous cathodes and possibly influence the tailoring of optimum electrode porosity and tortuosity.

**Acknowledgements**

This work was supported by the Toyota Research Institute of North America (TRINA). J.S and S.S contributed equally to this work.

**Figures and Tables**

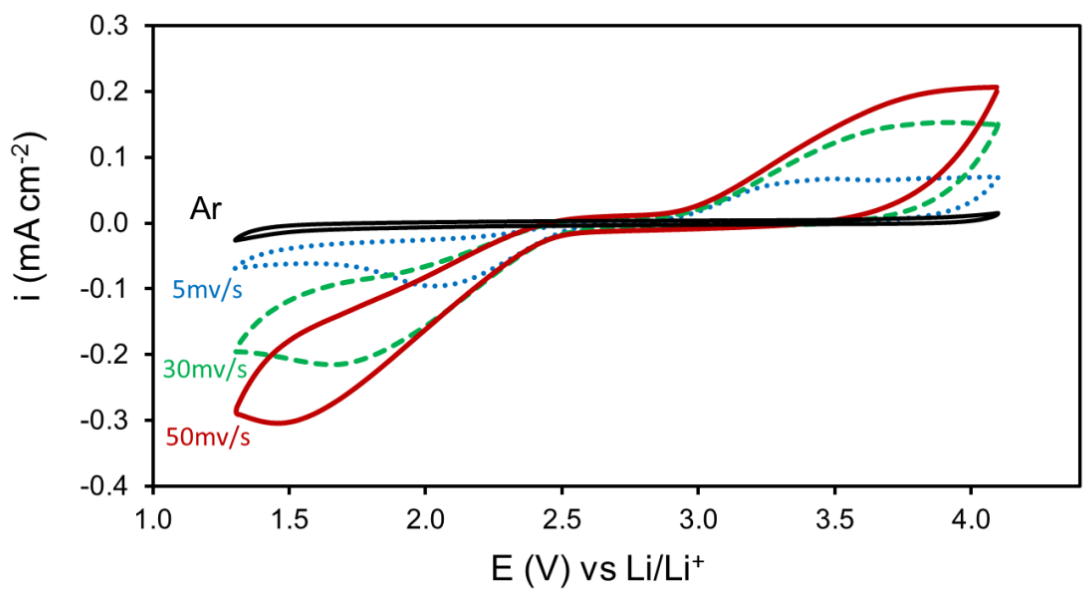

**Figure 1.** Cyclic voltammograms on a 1mm carbon black film (10% PTFE) deposited on a 0.196cm$^2$, roughened glassy carbon (GC) disk electrode in Ar and O$_2$ saturated 0.1 M LiTFSI in DME with varying scan rate.

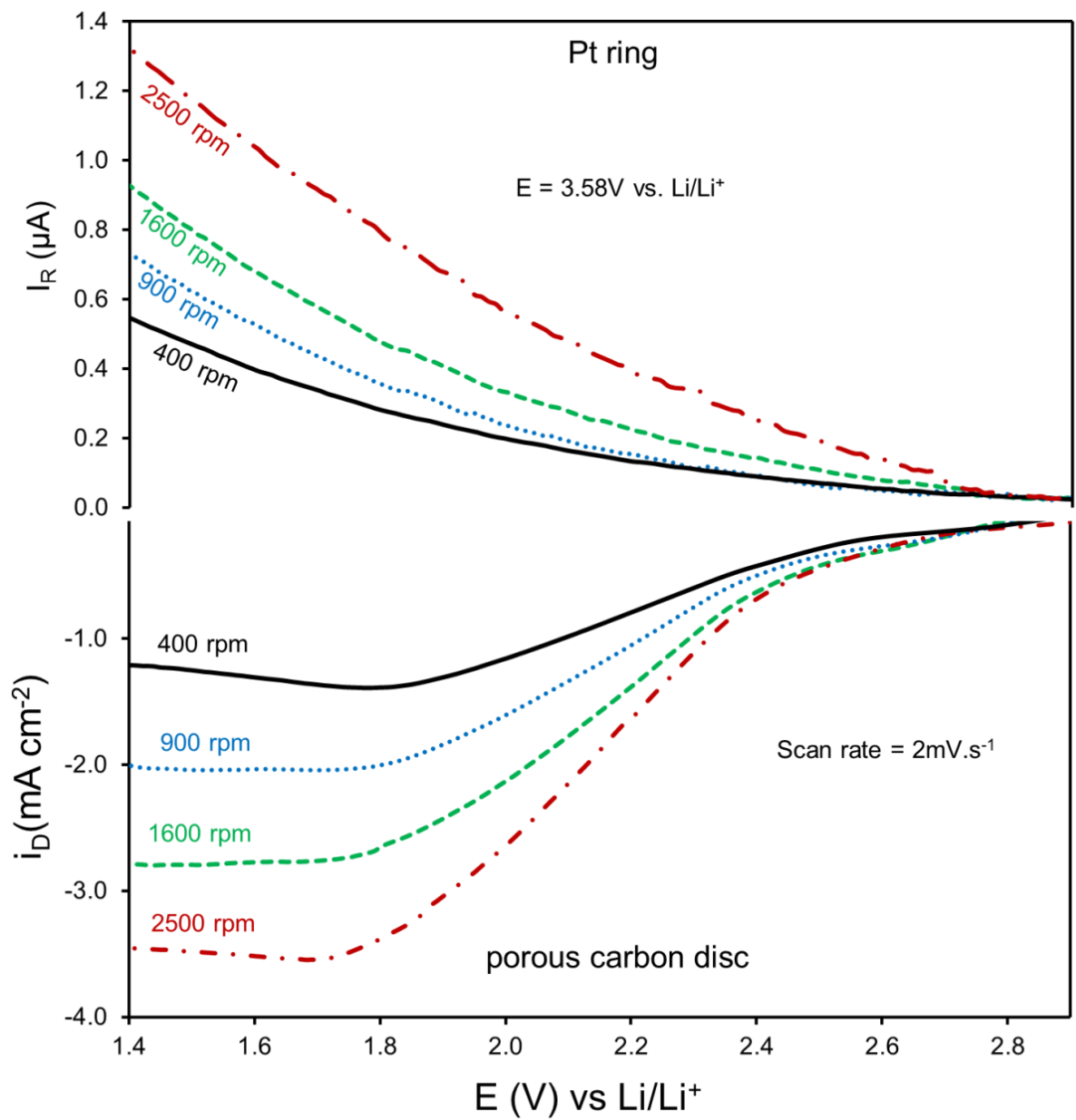

**Figure 2.** Linear sweep voltammograms on a rotating ring-disk electrode with a 1mm carbon black film (10% PTFE) deposited on a 0.196cm$^2$, roughened glassy carbon (GC) disk and platinum ring in O$_2$ saturated electrolyte. Disk currents were recorded at a scan rate of 2mVs$^{-1}$. Ring currents recorded with the Pt ring held at 3.08V vs Li/Li$^+$.

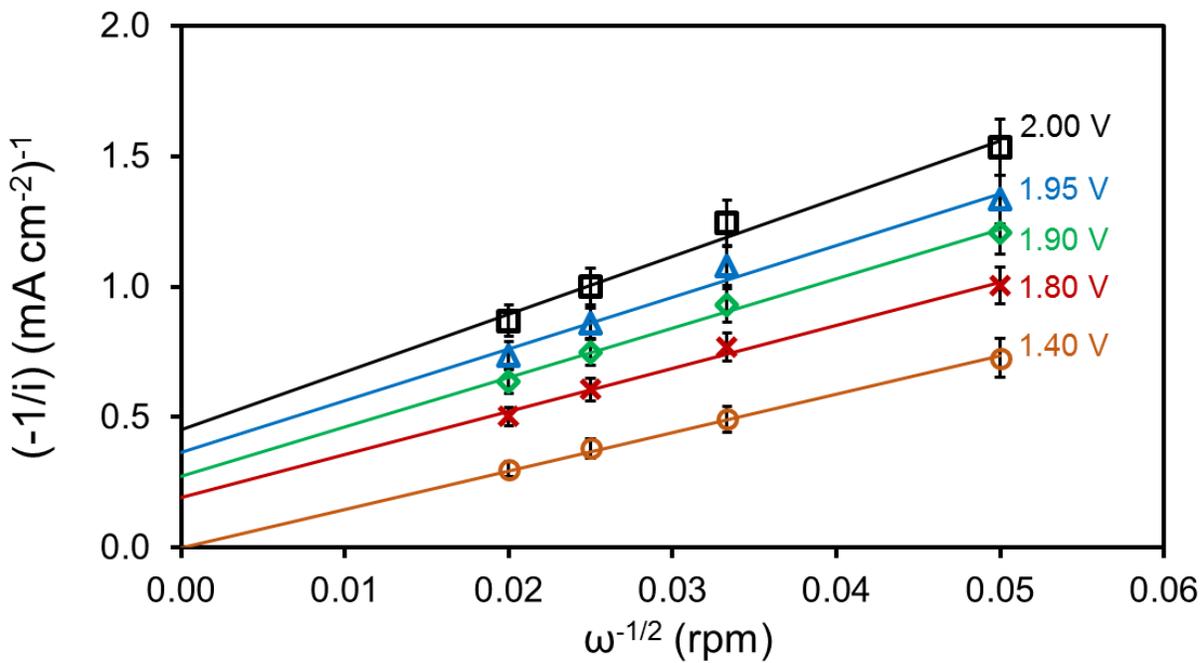

**Figure 3.** Koutecky-Levich ($-i^{-1}$ vs $\omega^{-1/2}$) plot for ORR on a 1mm carbon black film (10% PTFE) deposited on a 0.196cm$^2$, roughened glassy carbon (GC) disk electrode in 0.1M LiTFSI/DME electrolyte. Error bars depict 10% standard error.

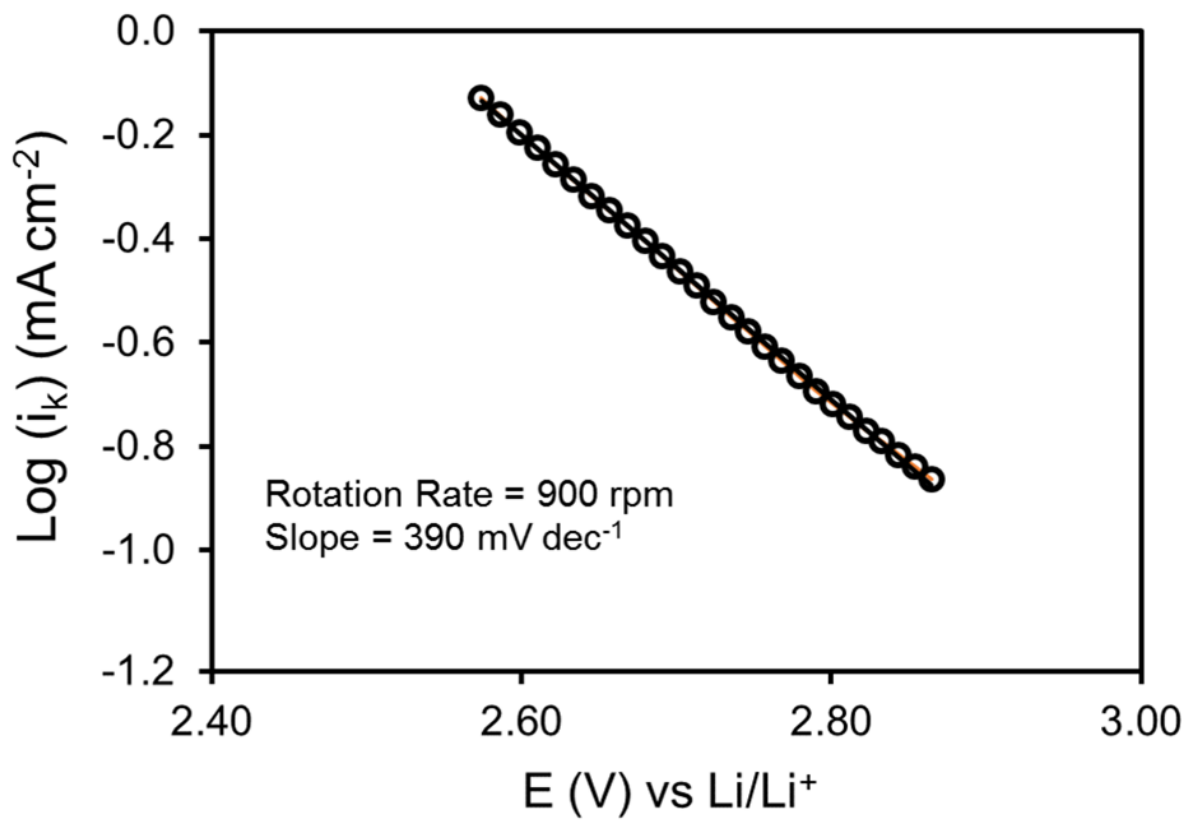

**Figure 4.** Tafel (log ($i_k$) vs E (V)) plot for ORR on a 1mm carbon black film (10% PTFE) deposited on a 0.196cm$^2$, roughened glassy carbon (GC) disk electrode in 0.1M LiTFSI/DME electrolyte.

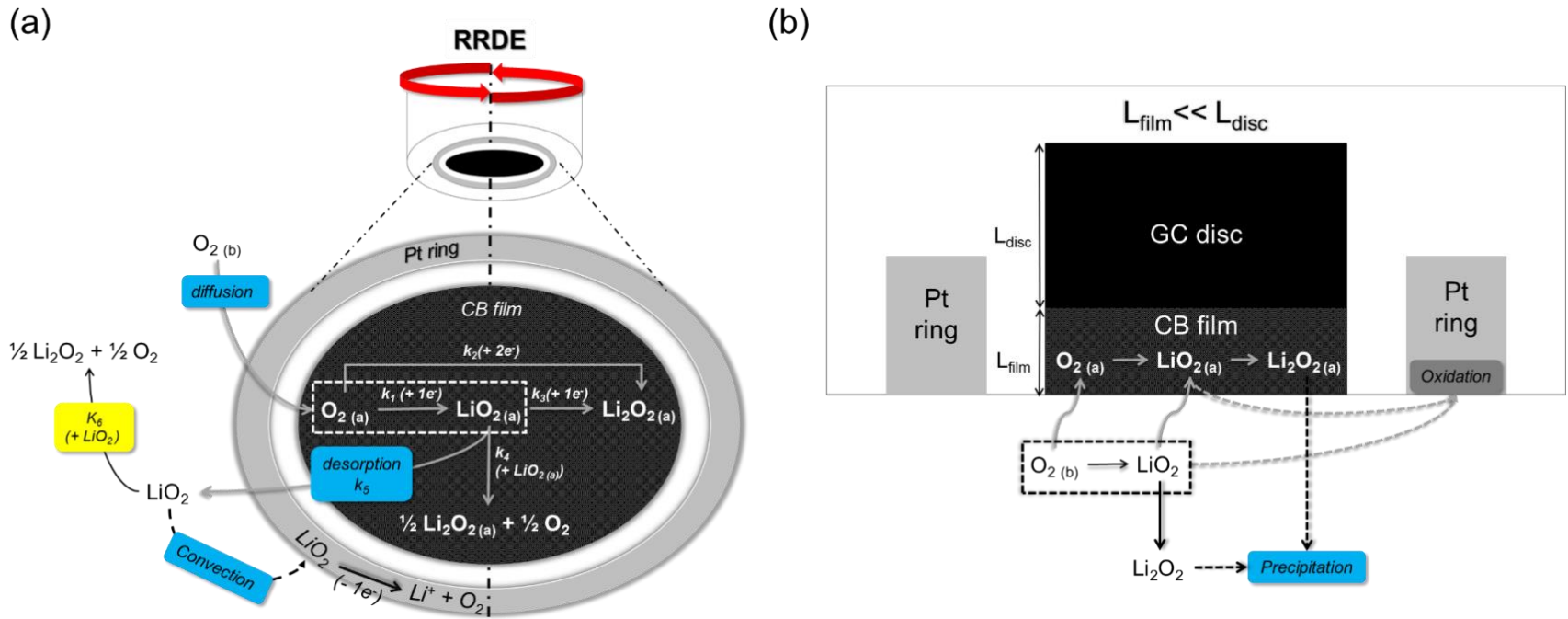

**Figure 5.** (a) Reaction scheme for the Oxygen reduction reaction on a Li-$O_2$ cell cathode and the kinetic model used with a Rotating ring-disk electrode (RRDE); (b) Cross-section view of the porous carbon black (CB) film disk electrode and the reactions therein (the reactions are illustrative and non-stoichiometric).

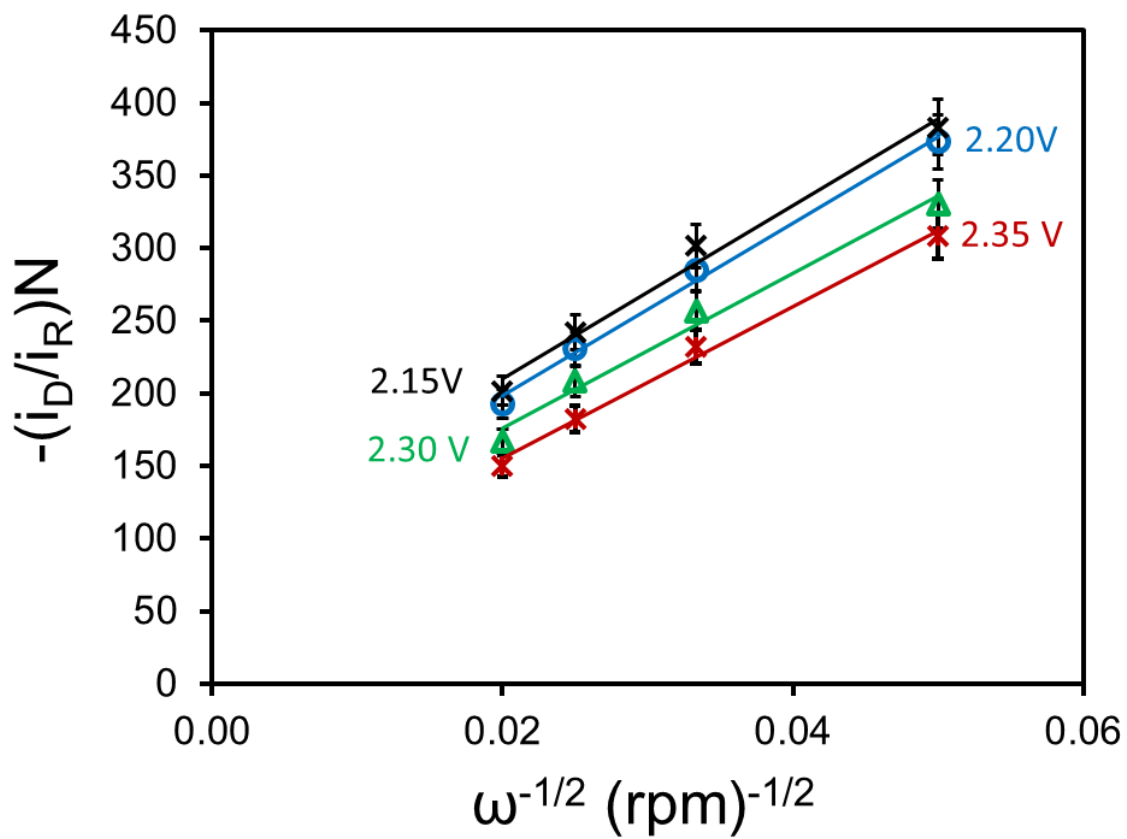

**Figure 6.** N($i_D/i_R$) vs $\omega^{-1/2}$ plots for ORR on a 1mm carbon black film (10% PTFE) deposited on a 0.196cm$^2$, roughened glassy carbon (GC) disk electrode in 0.1M LiTFSI/DME electrolyte.

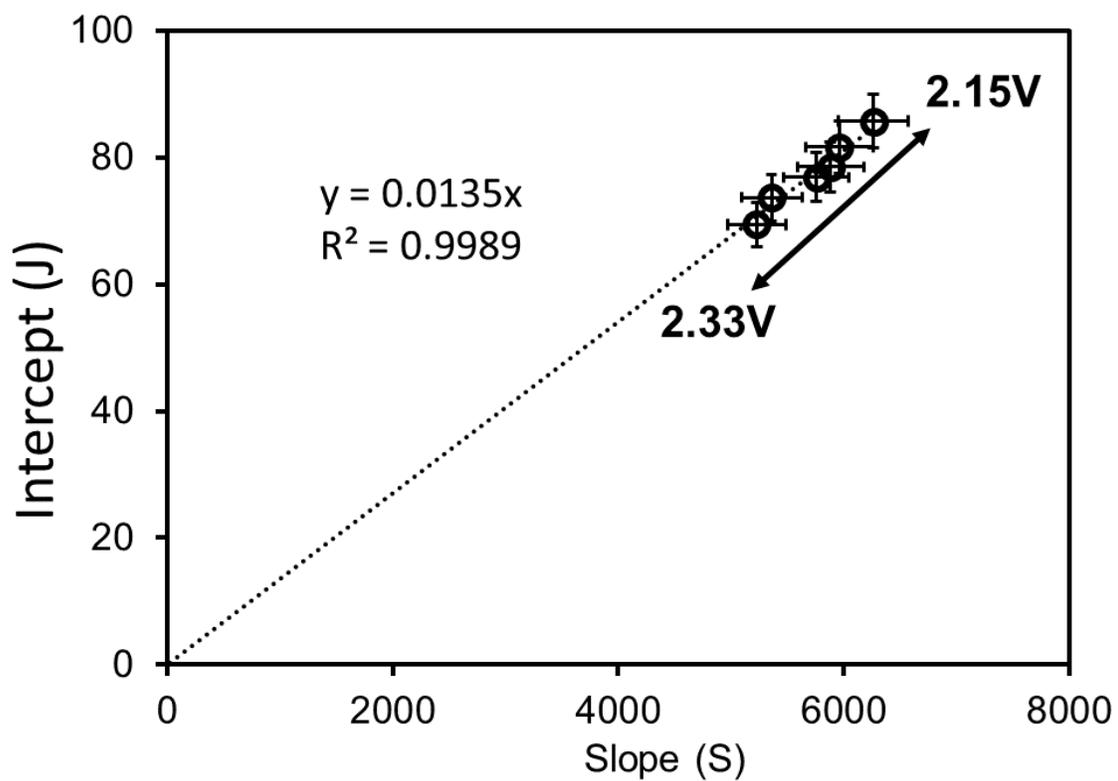

**Figure 7.** Slope(S) and intercept (J) plot from $N(i_D/i_R)$ vs $\omega^{-1/2}$ plots for ORR on a 1mm carbon black film (10% PTFE) deposited on a $0.196 cm^2$, roughened glassy carbon (GC) disk electrode in 0.1M LiTFSI/DME electrolyte.

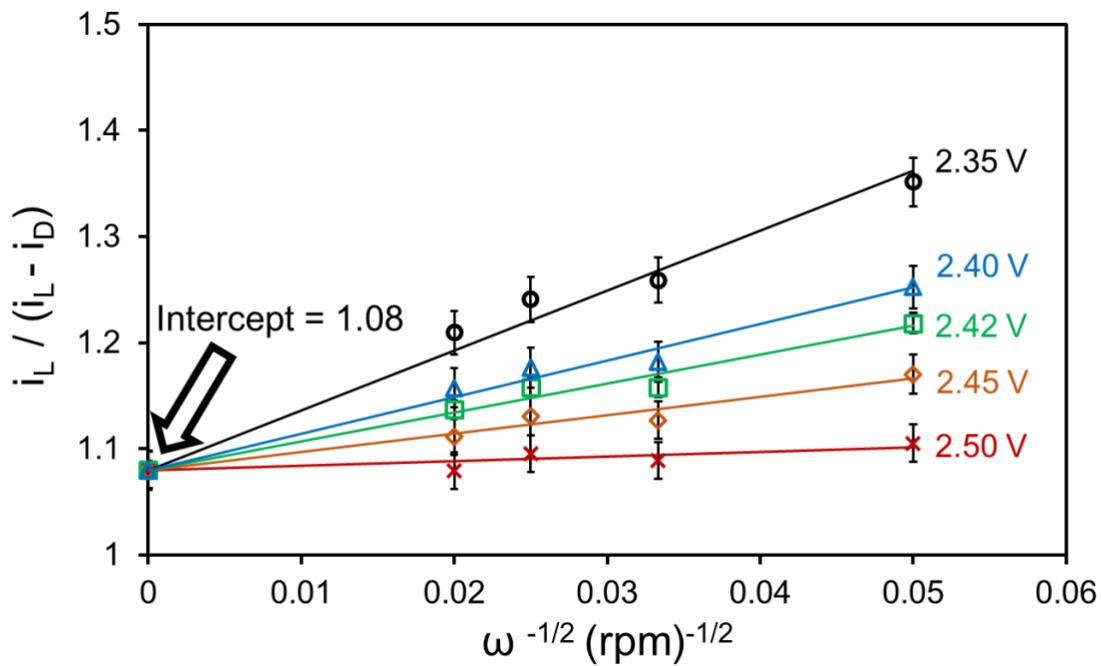

**Figure 8.** $i_L/(i_L-i_D)$ vs. $\omega^{-1/2}$ plots at different potentials for ORR on a 1mm carbon black film (10% PTFE) deposited on a 0.196cm$^2$, roughened glassy carbon (GC) disk electrode in 0.1M LiTFSI/DME electrolyte. Error bars depict 10% standard error.

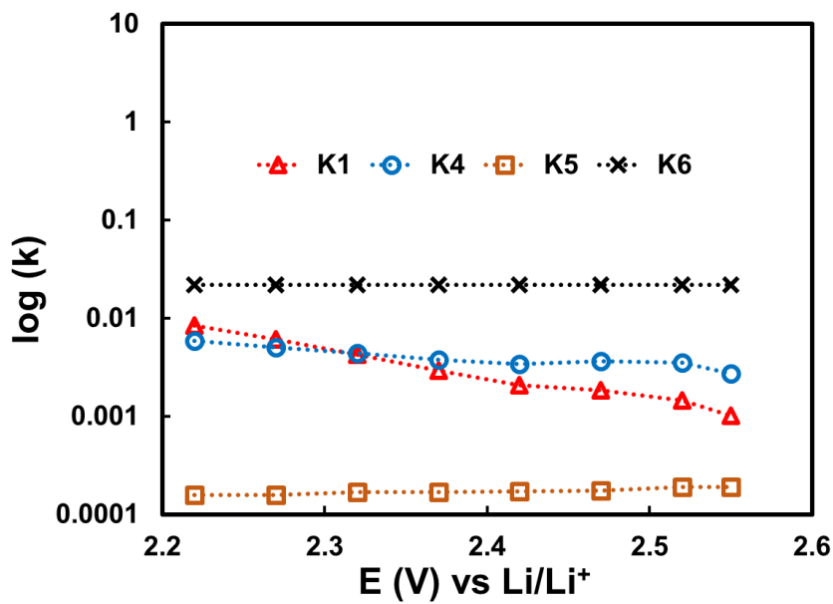

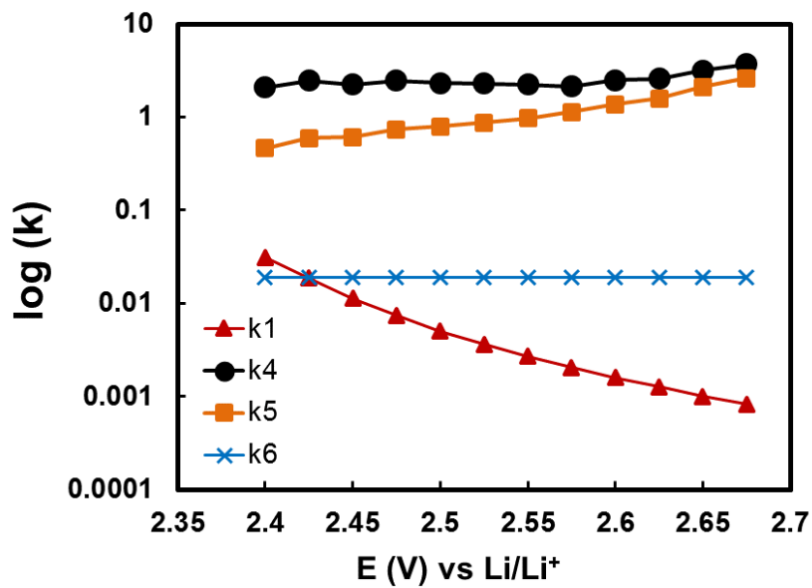

**Figure 9.** Calculated rate constants on **(a)** 1mm carbon black film (10% PTFE) deposited on a 0.196cm$^2$, roughened glassy carbon (GC); **(b)** smooth, planar glassy carbon. Data for (b) is taken from Sankarasubramanian et al. [22].

# Supporting Information

# Effect of cathode porosity on the Lithium air cell oxygen reduction reaction – a rotating ring-disk electrode investigation

Jeongwook Seo[a,*,#] Shrihari Sankarasubramanian[a,*,#,**], Nikhilendra Singh[b], Fuminori

Mizuno[b,c], Kensuke Takechi[b], and Jai Prakash[a,*]

[a] *Center for Electrochemical Science and Engineering, Department of Chemical and Biological Engineering, Illinois Institute of Technology, 10 West 33rd Street, Chicago, IL 60616, USA*

[b] *Toyota Research Institute of North America, 1555 Woodridge Avenue, Ann Arbor, Michigan 48105, USA*

**Table of contents:**

**1.**      **Calibration of the Pt pseudo-reference electrode**

---

[*] To whom correspondence may be addressed: E-mail address: jseo16@hawk.iit.edu (J.Seo), ssanka11@hawk.iit.edu (S, Sankarasubramanian), prakash@iit.edu (J.Prakash)

\# These authors contributed equally to this work.

[c] Present address: Battery Material Engineering & Research Division, Toyota Motor Corporation, Higashifuji Technical Center, 1200, Mishuku, Susono, Shizuoka, 410-1193 Japan

[**] Present address: Department of Energy, Environmental and Chemical Engineering, Washington University in St. Louis, 1 Brookings Dr, St. Louis, MO 63130, USA

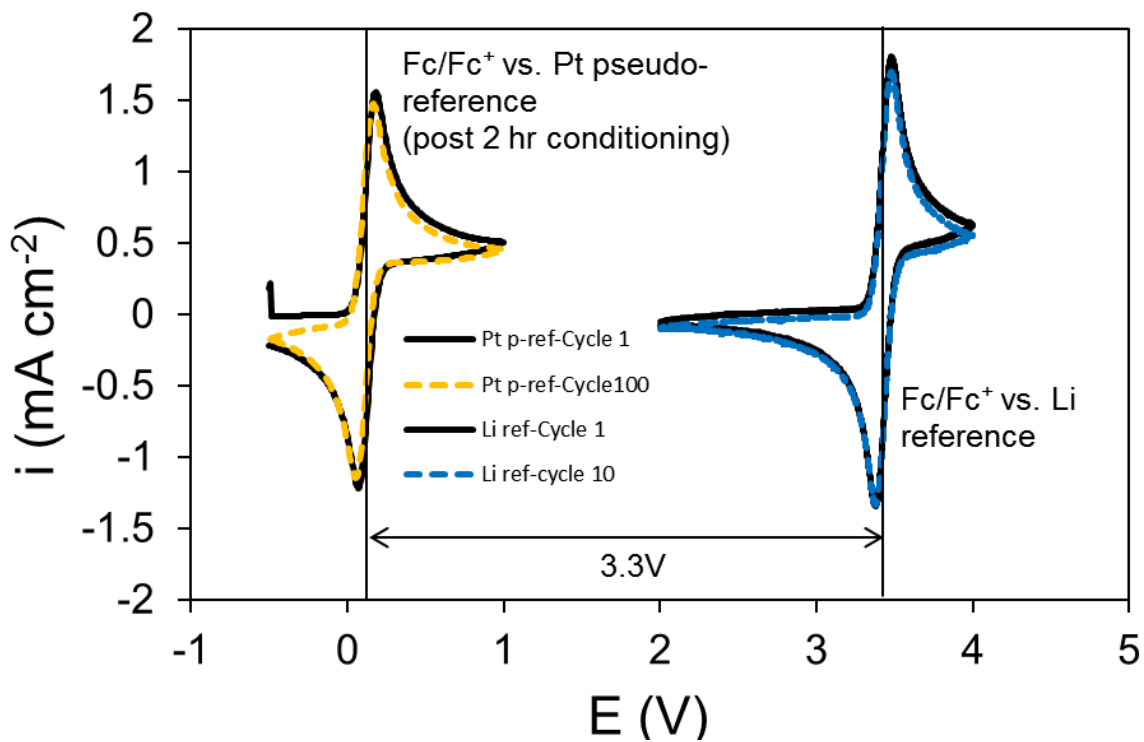

**Figure S1.** Calibration of the Pt pseudo-reference electrode against the Li/Li$^+$ reference electrode using the ferrocene/ferrocenium (Fc/Fc$^+$) couple as an internal standard. The electrolyte was 0.5M LiTFSI in DME with 10mM ferrocene added. Scan rate = 50mV.s$^{-1}$.

The Pt pseudo-reference was calibrated using the ferrocene/ferrocenium internal reference as recommended by IUPAC [1]. 0.5M LiTFSI in DME was prepared in an Ar-filled glovebox($H_2O$ and $O_2$ <0.5ppm) and 10mM ferrocene added to it. The cyclic voltammograms were measured in a 3-electrode setup with a glassy carbon working electrode, a fritted Pt spiral counter electrode, with a Li metal reference, and subsequently a Pt mesh pseudo-reference. Multiple consequtive cycles were carried out to establish repreoducibility and to characterize potential drift (if any). The results are shown in Figure S1. As is evident, the Pt pseudoreference is stable within the time scale of our experiments and is seen to be suitable as a reference electrode.